\documentclass[conference]{IEEEtran}
\IEEEoverridecommandlockouts
\usepackage{cite}
\usepackage{amsmath,amssymb,amsfonts}
\usepackage{algorithmic}
\usepackage{graphicx}
\usepackage{textcomp}
\usepackage{xcolor}
\usepackage{url}
\makeatletter
\g@addto@macro{\UrlBreaks}{\UrlOrds}
\makeatother
\usepackage{float}
\usepackage{listings}
\usepackage{flushend}

\colorlet{punct}{red!60!black}
\definecolor{background}{HTML}{EEEEEE}
\definecolor{delim}{RGB}{20,105,176}
\colorlet{numb}{magenta!60!black}

\lstdefinelanguage{json}{
    basicstyle=\normalfont\ttfamily,
    stepnumber=1,
    numbersep=8pt,
    showstringspaces=false,
    breaklines=true,
    frame=lines,
    literate=
     *{0}{{{\color{numb}0}}}{1}
      {1}{{{\color{numb}1}}}{1}
      {2}{{{\color{numb}2}}}{1}
      {3}{{{\color{numb}3}}}{1}
      {4}{{{\color{numb}4}}}{1}
      {5}{{{\color{numb}5}}}{1}
      {6}{{{\color{numb}6}}}{1}
      {7}{{{\color{numb}7}}}{1}
      {8}{{{\color{numb}8}}}{1}
      {9}{{{\color{numb}9}}}{1}
      {:}{{{\color{punct}{:}}}}{1}
      {,}{{{\color{punct}{,}}}}{1}
      {\{}{{{\color{delim}{\{}}}}{1}
      {\}}{{{\color{delim}{\}}}}}{1}
      {[}{{{\color{delim}{[}}}}{1}
      {]}{{{\color{delim}{]}}}}{1},
}

\newcommand{\reffig}[1]{Fig.~\ref{#1}}
\def\BibTeX{{\rm B\kern-.05em{\sc i\kern-.025em b}\kern-.08em
    T\kern-.1667em\lower.7ex\hbox{E}\kern-.125emX}}
\begin{document}

\title{Long Live The Image: Container-Native Data Persistence in Production
\thanks{This research is supported in part by Chilean National Research and Development Agency (ANID, Chile) under Grant
FONDECYT Iniciaci{\'o}n 11180905.}
}

\author{\IEEEauthorblockN{Zheng Li}
\IEEEauthorblockA{\textit{Department of Computer Science, University of Concepci{\'o}n} \\
Concepci{\'o}n, Chile \\
ORCID: 0000-0002-9704-7651}
}

\maketitle

\begin{abstract}
Containerization plays a crucial role in the de facto technology stack for implementing microservices architecture (each microservice has its own database in most cases). Nevertheless, there are still fierce debates on containerizing production databases, mainly due to the data persistence issues and concerns. Driven by a project of refactoring an Automated Machine Learning system, this research proposes the container-native data persistence as a conditional solution to running database containers in production. In essence, the proposed solution distinguishes the stateless data access (i.e.~reading) from the stateful data processing (i.e.~creating, updating, and deleting) in databases. A master database handles the stateful data processing and dumps database copies for building container images, while the database containers will keep stateless at runtime, based on the preloaded dump in the image. Although there are delays in the state/image update propagation, this solution is particularly suitable for the read-only, the eventual consistency, and the asynchronous processing scenarios. Moreover, with optimal tuning (e.g., disabling locking), the portability and performance gains of a read-only database container would outweigh the performance loss in accessing data across the underlying image layers.  
\end{abstract}


\section{Introduction}
As a lightweight virtualization technique, containerization plays a crucial role in the de facto technology stack for implementing microservices architecture and developing microservices systems \cite{Singh_2017}. Following the microservices architecture principles, each microservice is defined as an independent, self-contained, and single-purpose application service for a unique business capability.  Correspondingly, in a microservice-based system, each microservice usually has its own database with respect to a particular bounded context of the system, so as to reduce the software complexity and improve the system scalability \cite{Barbosa_2020}.

When creating container-based microservices, however, there are fierce debates on running the microservice's database component inside a container in the production environment, and the concerns about production database containers are mostly related to the data persistence. Given the ephemeral nature of containers at runtime, it is widely accepted that containers by design are neither to handle state nor to store data. As a result, the common practice of containerizing databases is to mount host or network directories (or files) as volumes to containers for saving and sharing data \cite{Krzywiec_2019}. Although some advocators expect to use this way to gain more scalability and flexibility in the database system, many arguments oppose this practice in production for its potential performance, reliability, agility, and data consistency issues. Instead, the opponents suggest outsourcing production databases directly to the well-managed database services in the public cloud.

Driven by the requirement of refactoring a private cloud-based Automated Machine Learning (AutoML) system, we still decided to explore the opportunity of fully-containerized microservices, in order to balance the workload between the application engineering team and the private cloud engineering team. Our ongoing efforts have led to a conditional solution to running database containers in production. By preloading data into a container image, this solution lets database containers support the stateless data access (i.e.~reading) only, while asking a master database to deal with the stateful data processing (i.e.~creating, updating, and deleting). By merging image-building activities with the master database's backup routine, this solution periodically rebuilds the container image and eventually syncs the new data to production.

Due to the intentionally decoupled stateless data access and stateful data processing, there are inevitably delays in the state/image update propagation from the master database to the containerized database instances. However, we argue that our solution is particularly applicable when the delay has little impact on real-world business values, for example in the read-only, the eventual consistency, and the asynchronous processing scenarios. More importantly, through this container-native data persistence approach, our solution can bring various benefits, such as:
\begin{itemize}
\item No host storage is required for running database containers, which avoids the frequent issue of dangling volumes due to the removal of containers.
\item No network storage is required for running database containers, which reduces the networking latency at least for the stateless data access.
\item No extra technique or tool is required for scaling and managing database containers, which decreases the complexity in container orchestration of the whole system.
\item No locking mechanism is required in the containerized database instances, which can help improve the database performance at least for the stateless data access.
\item No dedicated backup policy is required for the master database, as database dumps are naturally needed for building images and syncing updates to production.
\end{itemize}

This paper reports the initial results of the work-in-progress of this research. Section \ref{sec:meet} briefly summarizes the existing practices of container-related data persistence, and particularly highlights the debates on running database containers. Section \ref{sec:solution} describes our conditional solution to running database containers in production, together with some initial while positive validation result. Section \ref{sec:conclusion} draws conclusions and discusses the needs and directions of the future work.

\section{When Container Meets Data Persistence}
\label{sec:meet}
In general, persistence is ``the continuance of an effect after its cause is removed''. In the context of data, persistence is to use non-volatile storage to keep data for future use. Despite the various formats, data can persist in two main ways, namely (disk-based) files and databases. This research only focuses on databases when discussing container data persistence. 

\subsection{Running a Database inside a Container}
It has been widely accepted that containers are designed neither to handle state nor to store data. 
Although each container has a thin writable layer on top of a stack of image layers, the writable layer is considered volatile, because containers are disposable/ephemeral at runtime. Once a container is terminated, both its writable layer and the data stored on it are removed. As such, the de facto solution to running a database inside a container is essentially to separate the physical database files from the database management system (DBMS). The DBMS software is encapsulated into an image for running the database container, while the physical database files are persistently stored on a local or network drive that can be mounted as a volume when the container runs.

\subsection{Debates on Running a Database inside a Container}

Ideally, separating storage from computation can significantly improve databases' scalability and flexibility in the frontend capacity planning and the backend resource provisioning \cite{Rahman_2019}.   
In practice, however, the separation between DBMS software and physical database files will not only make the application performance decline but also make the error rates and latency increase, especially when the database files reside on the remote storage \cite{Bisson_2016,Visheratin_2020}. 

When it comes to containerizing databases, even if we constrain the database file location to be the host machine's hard disk, a containerized database will become less agile and less relocatable when it involves host-specific configurations \cite{Bisson_2016}, not to mention when it requires considerable disk space to store large amounts of data \cite{Rahman_2019}. More critically, the mechanism itself of mounting data volumes ``by punching a hole through the container'' can be unreliable \cite{Tobin_2016}.

Furthermore, there exists a dilemma about scaling database containers together with their volumes \cite{Krzywiec_2019}. On one hand, scaling out such database instances may not only generate unexpected  dangling volumes, but also result in data consistency issues once different data are stored in different volumes. On the other hand, if we try to avoid those issues by running a single database instance only, it will make little sense to containerize the database, as we will not take advantage of most powerful features of orchestration tools, e.g., automated container deployment, scaling, and management.   

Consequently, the compromise opinion in the community seems to be running database containers in the development and test/demo environments, while employing the fully managed cloud database services for production databases \cite{Heddings_2020}. 

\subsection{An Arguable Solution to Running Database Containers in Production}
\label{subsec:arguable}
To enable running database containers in production, an alternative solution claims to make data persist at the application level instead of at the infrastructure level, for example by taking advantage of the self-clustering feature of particular database products (e.g., Couchbase) \cite{Bisson_2016}. With this feature, a cluster of database instances can back-up and update each other automatically. When a new database instance joins the cluster, it also automatically replicates the data from the other instances. As such, even if we containerize the  database instances without using external volumes, this solution can still achieve data persistence by storing data inside the ephemeral containers, as long as there remain active database containers in the cluster all the time.

Nevertheless, this solution has been argued to be of high complexity even with respect to the basic database functionality, i.e.~keeping data alive and available \cite{Tobin_2016}. Furthermore, it is too good to be true that we never lose the entire cluster of database containers. In fact, this solution proposal also suggests that ``the primary instance is periodically backed up to an object store''. Thus, sophisticated fault-tolerance mechanisms (e.g., runtime cluster reconstruction) will still be needed if we implement this solution in production, no matter how big the cluster is.

\section{A Conditional Solution to Running Database Containers in Production}
\label{sec:solution}
This research was triggered by an AutoML project in collaboration with the local government. 
In this project, the infrastructure engineering team developed a private cloud and the application engineering team implemented a set of machine/deep learning (ML) workflows in the private cloud. When using this AutoML system, users can submit satellite maps (and make self-defined labels if training), and then selecting proper ML techniques to conduct real-world tasks (e.g., identifying available wood resources with geographic coordinates). Driven by the requirement of better scalability and maintainability, the current effort in this project is to use microservice technologies to refactor the AutoML system, before integrating new ML techniques. In particular, we proposed this lightweight and container-native data persistence solution for running database containers in suitable production scenarios.

\subsection{Solution Prototype}
\begin{figure}[htbp]
\centerline{\includegraphics[width=8cm,trim=40 567 315 78,clip]{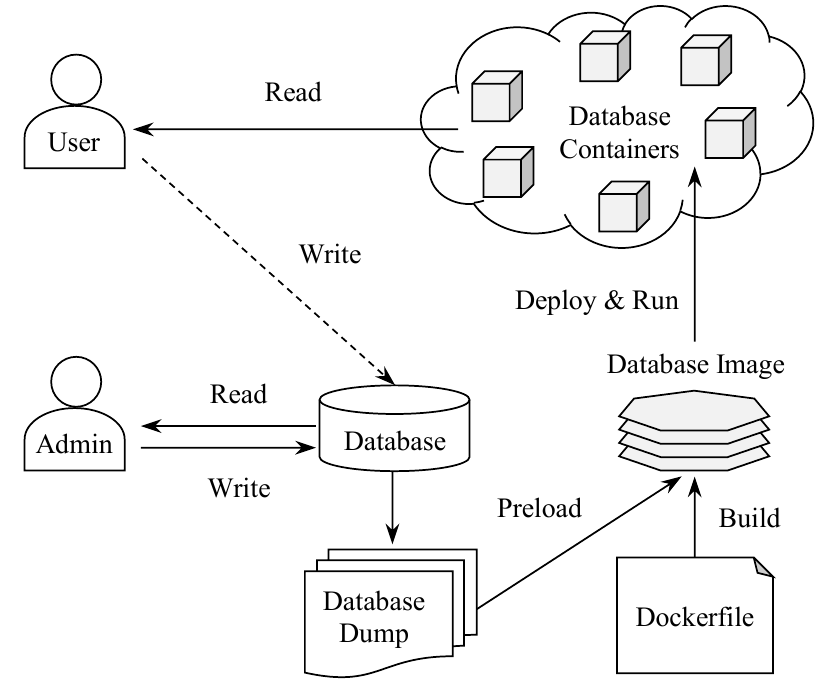}}
\caption{Data flow of container-native data persistence in production.}
\label{fig:solution}
\end{figure}

Unlike the strategy described in Section \ref{subsec:arguable} that stores data on the writable layer of containers, this research advocates container-native data persistence by injecting data when building images, as illustrated in \reffig{fig:solution}. Recall that an image is an immutable file that can be viewed as a static container template. The containers based on the same image will all be identical, and the data stored in the image will never change. Thus, by preloading data into a database image, we can completely rely on the scheduler and orchestration tools (e.g.,~Kubernetes) to take care of the database containers at runtime, without necessarily supplementing any fault-tolerance mechanism or even being aware of the containers' lifecycles. 

Due to the immutable nature of images, the running containers in our solution are essentially read-only database instances. When it comes to writing data into the database, the user's and/or system administrator's input will be redirected to a separate master database. The master database generates full database dumps periodically or on demand in the form of SQL statements. The generated database dumps can not only play a backup role, but also be used to update (by rebuilding) the database image. To enhance the time efficiency, the database image can be built offline and then be shared/deployed to the production environment for running database containers.

As for replacing the running database containers with new versions based on the new image, we propose to integrate the rolling update techniques \cite{Rossi_2020} into our solution to enable the zero-downtime replacement in production, which however is outside the focus of this current prototype. Overall, to address (at least relieve) the challenge of statefulness in microservice implementations, 
the essential argument in this research is to distinguish the stateless data access (i.e.~reading data only) from the stateful data processing (i.e.~creating, updating, and deleting data) in databases. This, on the other hand, leads to a couple of conditions of applying our solution, as discussed in the following subsection.

\subsection{Solution Conditions and Applicable Scenarios}
\label{subsec:discussion}
In brief, the solution conditions include: (1) It is possible to make the reading and writing transactions loosely coupled in one business logic; and (2) There is no negative impact on the real-world business value if the reading and writing transactions are made loosely coupled. To help clarify these conditions, we identify three scenarios that would be suitable for applying our solution.
\begin{itemize}
\item \textbf{The read-only scenario} is when users only need to obtain data from database. In the AutoML project, the read-only component is the API catalog of available ML techniques, which has been redesigned and implemented as a microservice. Although the API catalog will still need update after the ML technique library is modified, the update is maintained by the system administrator, while the users do not need (and are not allowed) to do anything except for browsing and choosing ML APIs. In fact, the similar use case widely exists in practice, for example the usage of online product catalog in e-commerce systems. Moreover, compared with the other business logic components, a developed catalog may require few updates, as the cataloged items are usually limited and fixed. Thus, this is the most suitable scenario for applying our solution.

\item \textbf{The eventual consistency scenario} is when users need to write to database, while they tolerate the delay in propagating the data changes to all the database instances. In the AutoML project, the satellite map labels created by users will not be merged into the map-label dataset until the next round of model training, which implies that the user data are made consistent eventually instead of immediately. Eventual consistency is also common in the real-world applications. Take a recommender system as an example, although users' input should always be included for future recommendation (e.g., based on a user-item incidence matrix), it is impossible to update the system with new data in real time, because the immediate update may freeze the whole system if there are many concurrent inputs. 
Therefore, it is possible to take advantage of users' delay tolerance in this case to rebuild images for including new data.

\item \textbf{The asynchronous processing scenario} is when users do not have to (or want to) stick to the data processing procedure, and they tolerate the delay in delivering data processing results. In the AutoML project, the users' ML tasks may experience not only the processing delay but also the queuing delay. Thus, the users can choose to receive email notifications or re-login the system later to find their task outcomes. A typical example of this scenario in our normal life is the flight booking via online travel agencies like Expedia. When travelers book flights online, although the reservations are made immediately, the travelers will have to wait up to days for the e-ticket numbers to appear. Therefore, we can slightly extend the data processing time for building and rebuilding images, without noticeably affecting the user experience.
\end{itemize}

\subsection{Initial Validation}
As mentioned previously, we have successfully applied our solution to the read-only scenario and run the container-native API catalog microservice in an AutoML project. To strengthen the validation, we also tested self-contained database containers by employing bigger-size datasets with more flexible configurations. For example, one of our trials built the database image by silently installing MySQL and then loading the open dataset of U.S. Geological Survey Geonames.\footnote{The source files for building the image is shared at \url{http://doi.org/10.5281/zenodo.4545788}} For the convenience of testing, we let the launched container automatically start the MySQL service and read out all the preloaded data. It should be noted that we emphasize preloading data when building images from scratch, which avoids the delay in loading data on startup of containers and also avoids the argument about committing container changes to new images. 

In addition to comparing the image sizes, we further verified whether the data were natively included or not by inspecting the launched containers. The inspection result confirms that no volume is mounted to the container based on our customized image. 

\begin{footnotesize}
\begin{lstlisting}[language=json,firstnumber=1,frame=single]
 "Mounts": [],
\end{lstlisting}
\end{footnotesize}

In contrast, by pulling the official MySQL image, the launched container will by default create and mount a volume within the directory of /var/lib/mysql on the host system, as shown below.

\begin{footnotesize}
\begin{lstlisting}[language=json,firstnumber=1,frame=single]
 "Mounts": [{
        "Type": "volume",
        "Name": "[Volume Name]",
        "Source": "/var/lib/docker/volumes/[Volume Name]/_data",
        "Destination": "/var/lib/mysql",
        "Driver": "local",
        "Mode": "",
        "RW": true,
        "Propagation": ""
 }],
\end{lstlisting}
\end{footnotesize}

At last, to verify the portability and migratability of our customized image, we copied the image to another Linux machine and a Windows 10 WSL 2 machine (via the pair of commands \texttt{docker save} and \texttt{docker load}). The launched containers successfully passed the test of data reading on both machines (with the Linux distribution Ubuntu 18.04 LTS). Although this data persistence solution is an anti-pattern against using faster volumes\footnote{\url{https://docs.docker.com/storage/volumes/}}, we believe that through optimal tuning (e.g., disabling locking), the portability and performance gains of read-only database containers may still outweigh their performance loss in accessing data across underlying image layers.

\section{Conclusions and Future Work}
\label{sec:conclusion}
When developing and deploying container-based microservices especially for cloud-native applications, the common practice is to employ the fully managed cloud database services rather than run database containers in the production environment. In the project of migrating a monolithic AutoML application to microservices architecture in a private cloud, however, we still decided to try production database containers, for balancing the workload between the application engineering team and the private cloud engineering team. To address the challenges and issues in containerizing production databases, we propose a container-native data persistence solution by decoupling stateless reading and stateful writing transactions, by preloading data into database images for stateless data access, and by using a master database to handle stateful data processing and to generate dumps periodically for rebuilding images. Our theoretical analysis and initial validation advocate that this conditional solution may particularly be suitable for the read-only, the eventual consistency, and the asynchronous processing scenarios.

Our future work will keep engineering the application and performance of this solution. As part of the validation work, we have verified the effectiveness and efficiency of our solution in the read-only scenario. As for the other scenarios, it will be necessary to optimize the trade-off between the delay and frequency of the update propagation. Furthermore, compared with the host-volume and network-volume solutions to container data persistence, although our solution avoids the networking latency for stateless data access, there might be some performance loss without using host volumes. Therefore, it will be valuable to empirically investigate and minimize the performance overhead of reading preloaded data from container images.


\bibliographystyle{IEEEtran}
\bibliography{ICSArefShort}

\end{document}